\documentclass{aa}  
\usepackage{graphicx}
\usepackage{txfonts}

\newcommand{\dg}{$^\circ$ } 
\newcommand{\dgg}{$^\circ$} 

\begin{document} 

   \title{In situ multi-spacecraft and remote imaging observations of the first CME detected by Solar Orbiter and BepiColombo}

   \author{E.E.~Davies \inst{1}
          \and C.~M\"ostl \inst{2,3}
          \and M.J.~Owens \inst{4}
          \and A.J.~Weiss \inst{2,3,5}
          \and T.~Amerstorfer \inst{2}
          \and J.~Hinterreiter \inst{2,5}
          \and M.~Bauer \inst{2}
          \and R.L.~Bailey \inst{6}
          \and M.A.~Reiss \inst{2,3}
          \and R.J.~Forsyth \inst{1}
          \and T.S.~Horbury \inst{1}
          \and H.~O'Brien \inst{1}
          \and V.~Evans \inst{1}
          \and V.~Angelini \inst{1}
          \and D.~Heyner \inst{7}
          \and I.~Richter \inst{7}
          \and H-U.~Auster \inst{7}
          \and W.~Magnes \inst{2}
          \and W.~Baumjohann \inst{2}
          \and D.~Fischer \inst{2}
          \and D.~Barnes \inst{8}
          \and J.A.~Davies \inst{8}
          \and R.A.~Harrison \inst{8}
          }

   \institute{Department of Physics, Imperial College London, London, UK \\
              \email{emma.davies12@imperial.ac.uk}
         \and Space Research Institute, Austrian Academy of Sciences, Graz, Austria
         \and Institute of Geodesy, Graz University of Technology, Graz, Austria
         \and Space and Atmospheric Electricity Group, Department of Meteorology, University of Reading, Reading, UK
         \and Institute of Physics, University of Graz, Graz, Austria
         \and Conrad Observatory, Zentralanstalt f\"ur Meteorologie und Geodynamik, Vienna, Austria
         \and Technical University of Braunschweig, Braunschweig, Germany
         \and STFC-RAL Space, Didcot, UK
             }

   \date {Received: 10 December 2020 | Revised: 1 February 2021 | Accepted: 20 February 2021}


  \abstract
   {On 2020 April 19 a coronal mass ejection (CME) was detected in situ by Solar Orbiter at a heliocentric distance of about 0.8~AU. The CME was later observed in situ on April 20 by the Wind and BepiColombo spacecraft whilst BepiColombo was located very close to Earth. This CME presents a good opportunity for a triple radial alignment study, as the spacecraft were separated by less than 5\dg in longitude. The source of the CME, which was launched on April 15, was an almost entirely isolated streamer blowout. The Solar Terrestrial Relations Observatory (STEREO)-A spacecraft observed the event remotely from -75.1\dg longitude, which is an exceptionally well suited viewpoint for heliospheric imaging of an Earth directed CME.}
   {The configuration of the four spacecraft has provided an exceptionally clean link between remote imaging and in situ observations of the CME. We have used the in situ observations of the CME at Solar Orbiter, Wind, and BepiColombo and the remote observations of the CME at STEREO-A to determine the global shape of the CME and its evolution as it propagated through the inner heliosphere.}
   {We used three magnetic flux rope models that are based on different assumptions about  the  flux  rope  morphology to interpret the large-scale structure of the interplanetary CME (ICME). The 3DCORE model assumes an elliptical cross-section with a fixed aspect-ratio calculated by using the STEREO Heliospheric Imager (HI) observations as a constraint. The other  two  models  are  variants  of  the  kinematically-distorted flux rope (KFR) technique, where two flux rope cross-sections are considered: one in  a  uniform  solar  wind and another in a solar-minimum-like structured solar wind. Analysis of CME evolution has been complemented by the use of (1) the ELEvoHI model to compare predicted CME arrival times and confirm the connection between the imaging and in situ observations, and (2) the PREDSTORM model, which provides an estimate of the Dst index at Earth using Solar Orbiter magnetometer data as if it were a real--time upstream solar wind monitor.}
   {A clear flattening of the CME cross-section has been observed by STEREO-A, and further confirmed by comparing profiles of the flux rope models to the in situ data, where the distorted flux rope cross-section qualitatively agrees most with in situ observations of the magnetic field at Solar Orbiter. Comparing in situ observations of the magnetic field between spacecraft, we find that the dependence of the maximum (mean) magnetic field strength decreases with heliocentric distance as $r^{-1.24 \pm 0.50}$ ($r^{-1.12 \pm 0.14}$), which is in disagreement with previous studies. Further assessment of the axial and poloidal magnetic field strength dependencies suggests that the expansion of the CME is likely neither self-similar nor cylindrically symmetric.}
   {}


   \titlerunning{The first CME detected by Solar Orbiter and BepiColombo}

   \maketitle

\section{Introduction} \label{sec:intro}

Coronal mass ejections (CMEs) are large-scale expulsions of plasma and magnetic fields that are expelled from the solar atmosphere. The simplest configuration of a CME observed in visible-light images has a three-part structure: a bright high-density leading loop enclosing a dark low-density cavity and a bright core \citep[e.g.][]{illing1985observation, vourlidas2013many}, where the cavity is thought to be a magnetic flux rope structure \citep{dere1999lasco, wood1999comparison}. 

In situ measurements of interplanetary CMEs (ICMEs) often display signatures of magnetic flux rope structures \citep{cane2003interplanetary}, likely the interplanetary manifestations of the magnetic flux rope associated with the dark cavity observed remotely \citep{burlaga1982}. ICMEs are the main drivers of severe space weather at Earth \citep[e.g.][]{kilpua2017goeffective} and therefore their evolution is of great interest in space weather modelling. To better understand ICME evolution in situ, it is useful to track signatures of specific ICMEs over large heliocentric distances whilst spacecraft are close to radial alignment \citep[e.g.][]{burlaga1981magnetic}. Studies of ICMEs observed by spacecraft separated by less than 5\dg in longitude and separated by more than ~0.2 AU in heliocentric distance are relatively rare but provide valuable insight into the properties, rotation, expansion, and interaction with other features of the solar wind environment as the ICME propagates \citep[e.g.][]{good2015radial, good2018correlation, winslow2016longitudinal,  kilpua2019, lugaz2019evolution, davies2020radial}.

The link between CMEs and ICMEs and their global structure and kinematics has previously been studied by using remote observations made by the Solar Terrestrial Relation Observatory \citep[STEREO;][]{kaiser2008stereo} in combination with in situ observations \citep[e.g.][]{davis2009stereoscopic, mostl2009linking, wood2009empirical, liu2010reconstructing, rouillard2010white}. The global structure of CMEs is a complex issue affected by the interaction of the CME with the solar wind and other CMEs as it propagates. The kinematic distortion of the large-scale structure of the CME in a latitudinally structured solar wind leads to a flattening or `pancaking' of the flux rope cross-section \citep{odstrcil1999distortion, riley_kinematic_2004, owens_magnetic_2006, owens2017coronal}, identified in remote observations \citep[e.g.][]{savani2010observational}. 

Understanding the evolution of CMEs through the inner heliosphere is one of the objectives of the Solar Orbiter mission \citep{mueller2013solar, mueller2020solar}, which launched on 2020 February 10. Solar Orbiter provides an opportunity to study CMEs both remotely and in situ, at closer heliocentric distances to the Sun (0.28~AU perihelion) and higher latitudes out of the ecliptic (up to 33\dgg). In this study, we present the magnetic field \citep[MAG;][]{horbury2020mag} observations of the first CME measured in situ at Solar Orbiter on 2020 April 19 whilst Solar Orbiter was in close radial alignment with both the Wind and BepiColombo \citep{benkhoff2010bepicolombo} spacecraft. During this period of time, BepiColombo was completing a flyby of Earth whilst on its cruise phase to Mercury, where the mission plans to investigate the interior, surface, exosphere, and magnetosphere to better understand the origin and evolution of the planet. The data are not public so far but available by request from the PI D. Heyner.

Solar Orbiter Energetic Particle Detector \citep[EPD;][]{rodriguez2020epd} observations of the Forbush decrease associated with the same CME are presented by \citet[in revision at A\&A,][]{vonforstner2020radial}. STEREO-A was well positioned to observe the CME remotely using Heliospheric Imager \citep[HI;][]{eyles2009heliospheric} data, providing an exceptional link to the in situ measurements to study the dynamical evolution and propagation of the CME. Section 2 presents the spacecraft observations and the flux rope models used to investigate the properties and large-scale structure of the CME as it propagates, the results of which are presented in Section 3 where both the radial and non-radial evolution of the CME has been assessed. Additionally, a modelling of the Dst index is demonstrated by using Solar Orbiter magnetometer data as if they were used as a real--time upstream solar wind monitor.

\section{Spacecraft data and methods}\label{sec:data}

\subsection{In situ observations} \label{sec:in-situ}

\begin{table}
\caption{ICME properties at Solar Orbiter, Wind, and BepiColombo, for 2020 April 19-21. ICME times $t$ are given, where the subscripts stand for (1) shock arrival time, (2) flux rope start time, (3) flux rope end time, (4) unperturbed magnetic flux rope (UMFR) start, (5) UMFR rope end. Also shown are the duration $\Delta$ for the full ICME, the sheath, the magnetic obstacle (MO) and the UMFR, and the mean total magnetic field in those 4 intervals. The spacecraft heliocentric distance $R$ is given at 3 different times, corresponding to the subscripts 1-3 above, as well as the HEEQ longitude $lon$ and HEEQ latitude $lat$.}             
\label{table:1}      
\centering                          
\begin{tabular}{c c c c c}        
\hline\hline                 
 & unit & Solar Orbiter  &  Wind  & BepiColombo \\     
\hline                        
   $t_1$  & UT        & Apr 19 05:06 &  Apr 20 01:34 &  Apr 20 03:09\\ 
   $t_2$ & UT        & Apr 19 08:59 &  Apr 20 07:56 & Apr 20 08:05\\
   $t_3$  & UT        & Apr 20 09:15 &  Apr 21 11:18 & Apr 21 10:08 \\
   $t_{4}$  & UT & Apr 19 13:56 &  Apr 20 11:28 & Apr 20 14:03 \\
   $t_{5}$  & UT & Apr 20 04:36 &  Apr 21 02:10 & Apr 21 03:18 \\
   $\Delta_{ICME}$ & h & 28.15   &  33.73  & 30.98 \\
   $\Delta_{sheath}$ & h & 3.88 & 6.37  & 4.93 \\
   $\Delta_{MO}$ & h &   24.27  & 27.37 & 26.05 \\
   $\Delta_{UMFR}$ & h &  14.67  & 14.70 & 13.25 \\
   $B_{ICME}  $ & nT &   17.2 & 12.3  &  12.7 \\
   $B_{sheath}$ & nT &   9.9  &  6.8  &  7.2 \\
   $B_{MO}$     & nT &   18.4 &  13.6 & 13.7 \\
   $B_{UMFR}$     & nT &   19.2 &  14.9 & 14.6 \\
   $R_1$ & AU          & 0.809 &  0.996  & 1.011\\
   $R_2$ & AU            & 0.808 &  0.996  & 1.011\\ 
   $R_3$ & AU               & 0.802 &  0.996  & 1.011\\ 
   $lon_1$& deg      & -4.02 &  0.19   & -1.25\\
   $lon_2$& deg        & -3.98 &  0.18   & -1.27\\ 
   $lon_3$& deg        & -3.70 &  0.18   & -1.41\\ 
   $lat_1$& deg      & -3.94  &  -5.15  & -5.53\\
   $lat_2$& deg        & -3.92 &  -5.13  & -5.52\\ 
   $lat_3$& deg          & -3.80 &  -5.03  & -5.47\\ 
\hline                                   
\end{tabular}
\end{table}

\begin{figure*}[h]
\includegraphics[width=\linewidth]{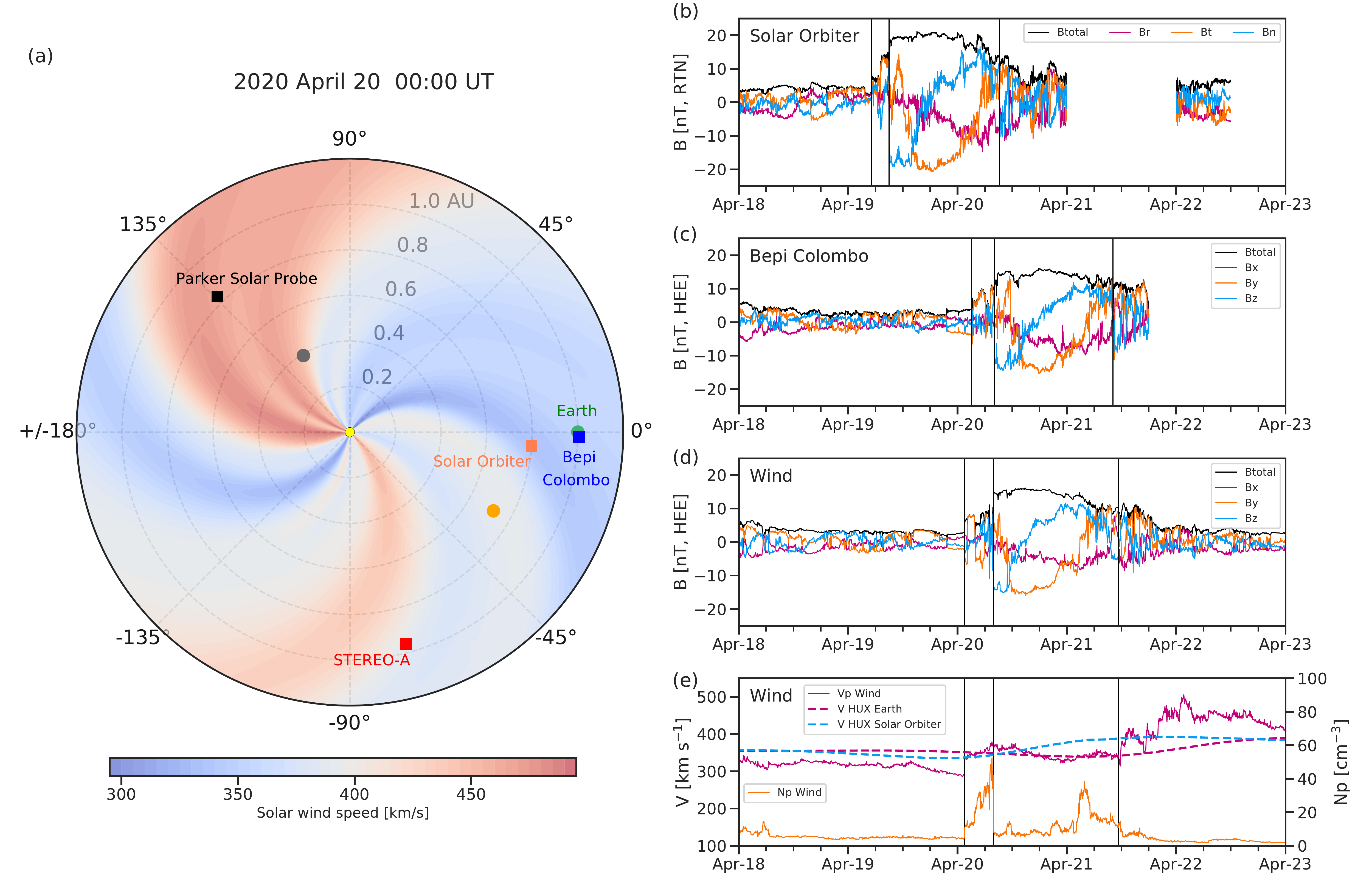}
\caption{Overview of spacecraft positions and in situ solar wind data. (a) WSA/HUX ambient solar wind speed in the ecliptic plane and spacecraft positions on 2020 April 20 00:00 UT in Heliocentric Earth Equatorial (HEEQ) coordinates. (b) Solar Orbiter magnetic field components in Radial-Tangential-Normal (RTN) coordinates, (c) BepiColombo magnetic field data in Heliocentric Earth Ecliptic coordinates (HEE), (d) Wind magnetic field data in HEE coordinates, and (e) the Wind proton speed and density (solid lines) and the WSA/HUX solar wind speed at Earth and Solar Orbiter (dashed lines). Vertical lines indicate from left to right the shock arrival time and the beginning and end times of the flux rope, determined visually. 
   \label{fig:1_overview} }
\end{figure*}

\textbf{Figure \ref{fig:1_overview}} gives an overview of spacecraft locations and in situ magnetic field and plasma data, taken by the Solar Orbiter magnetometer \citep[MAG;][]{horbury2020mag}, the magnetometer (MAG) on the Mercury Planetary Orbiter (MPO) of the BepiColombo spacecraft  \citep[MPO-MAG;][]{glassmeier2010fluxgate, heyner2020bepi}, and the Magnetic Field Investigation \citep[MFI;][]{lepping1995} and Solar Wind Experiment \citep[SWE;][]{ogilvie1995} of Wind. 

\textbf{Table~\ref{table:1}} summarises the arrival times of the shock, the start of the magnetic obstacle, which in this case is a clean flux rope with a rotating magnetic field, the end of the magnetic obstacle, the spacecraft position at those times and basic ICME parameters for the sheath, magnetic obstacle and the full ICME interval, consisting of the sheath and the magnetic obstacle \citep{rouillard2011} combined. The Solar Wind Analyzer \citep[SWA;][]{owens_swa_2020} instrument onboard Solar Orbiter was not yet fully commissioned during the time period the CME occurred, and therefore solar wind plasma data were only available at Wind, where it arrived with a low mean ICME speed of 346~km~s$^{-1}$. The selected boundaries will be discussed in more detail in the results section.

At the ICME shock arrival time on 2020 April 19 05:06 UT, Solar Orbiter was positioned at a heliocentric distance of 0.809~AU, and at $-4.02$\dg longitude and $-3.94$\dg latitude in Heliocentric Earth Equatorial (HEEQ) coordinates. Wind was near the Sun--Earth L1 point at 0.996 AU, 0.19\dg longitude, and -5.15\dg latitude at the shock arrival time on April 20 01:34 UT, about 20.5 hours after the shock arrived at Solar Orbiter. The shock then reached BepiColombo on April 20 03:09 UT, about 1.5 hours after it impacted Wind, when BepiColombo was located at 1.011 AU, $-1.25$\dg longitude and $-5.53$\dg latitude, still near Earth a few days after a flyby event. Additionally, STEREO-A was imaging the event from -75.1\dg longitude, which is an exceptionally well suited viewpoint for heliospheric imaging of an Earth directed CME. 

 \textbf{Figure \ref{fig:1_overview}a} shows the ambient solar wind conditions in the ecliptic plane on 2020 April 20 00:00 UT calculated with the Wang-Sheeley-Arge/Heliospheric Upwind eXtrapolation (WSA/HUX) model~\citep{reiss19,reiss20}. The WSA/HUX model is based on magnetic maps of the photospheric field from the Global Oscillation Network Group (GONG) provided by the National Solar Observatory (NSO). Specifically, we use the magnetic maps as an inner boundary condition to the Potential Field Source Surface~\citep[PFSS;][]{altschuler69} and the Schatten current sheet model~\citep[SCS;][]{schatten71} to compute the large-scale coronal magnetic field. Using the global magnetic field topology, we specify the solar wind conditions near the Sun based on the established Wang-Sheeley-Arge relation~\citep[WSA;][]{arge03}. To evolve the solar wind conditions from near the Sun to the Earth, we use the Heliospheric Upwind eXtrapolation model~\citep[HUX;][]{riley11b, owens2017probabilistic, reiss20} which simplifies the fluid momentum equation as much as possible. \textbf{Figure~\ref{fig:1_overview}a} shows the computed large-scale solar wind conditions in interplanetary space. The time interval under scrutiny reveals typical solar wind conditions during solar minimum with extended polar coronal holes and streamers confined within a narrow latitude band near the equator. Our model results show that a higher speed stream in the ambient solar wind is expected to arrive right after the ICME impacts Earth, and indeed it is seen in the solar wind speed time series at Wind (\textbf{Figure~\ref{fig:1_overview}e}) that this higher speed stream interacts with the rear of the ICME when it passes 1~AU. It is also seen that a rise in the HUX model speed matches to a good degree the observed rise in the in situ speed at Wind. 

Looking at the magnetic field data in \textbf{Figure \ref{fig:1_overview}b-d}, we can see that when each of the three spacecraft are in the magnetic obstacle (between the second and third vertical lines), the normal component $B_n$ or $B_z$ of the magnetic field is bipolar and turns from negative to positive, and the $B_t$ or $B_y$ component is mostly unipolar and negative, which in the HEE (used for the BepiColombo and Wind data) and Radial-Tangential-Normal (RTN) coordinate systems (used for Solar Orbiter) points towards solar east (note that differences in the axis directions are small between HEE and RTN). The unipolar component $-B_t$ or $-B_y$ (east) is also the direction of the flux rope axial field. This field rotation corresponds to a left-handed south-east-north (SEN) type of flux rope \citep{bothmer1998,mulligan1998}, which is of left-handed chirality and has a low inclination of the flux rope axial field to the solar equatorial plane.

The event is essentially a triple radial ICME lineup event observed by Solar Orbiter, Wind, and BepiColombo. The spacecraft separation between Solar Orbiter and Wind of 0.19~AU makes it possible to study the radial evolution of the ICME flux rope, as the difference in longitude from Wind of 4.2\dg is well below the estimated full longitudinal extent of ICME flux ropes of 60\dg \citep[e.g.][]{bothmer1998,good2016}. BepiColombo, with a 1.3\dg difference in longitude and 0.015~AU in radial distance from Wind, is too close to Wind to see significant differences in the large-scale shape or structure of the flux rope \citep{lugaz2018}. However, this does help in setting flux rope boundaries for Solar Orbiter and Wind to identify its coherent part, and is very well suited for multi-point studies of small scale features of the shock and sheath \citep[][in prep.]{kilpua2020energetic}.

\subsection{Heliospheric imaging} \label{sec:HI}

\begin{figure*}[h]
\includegraphics[width=\linewidth]{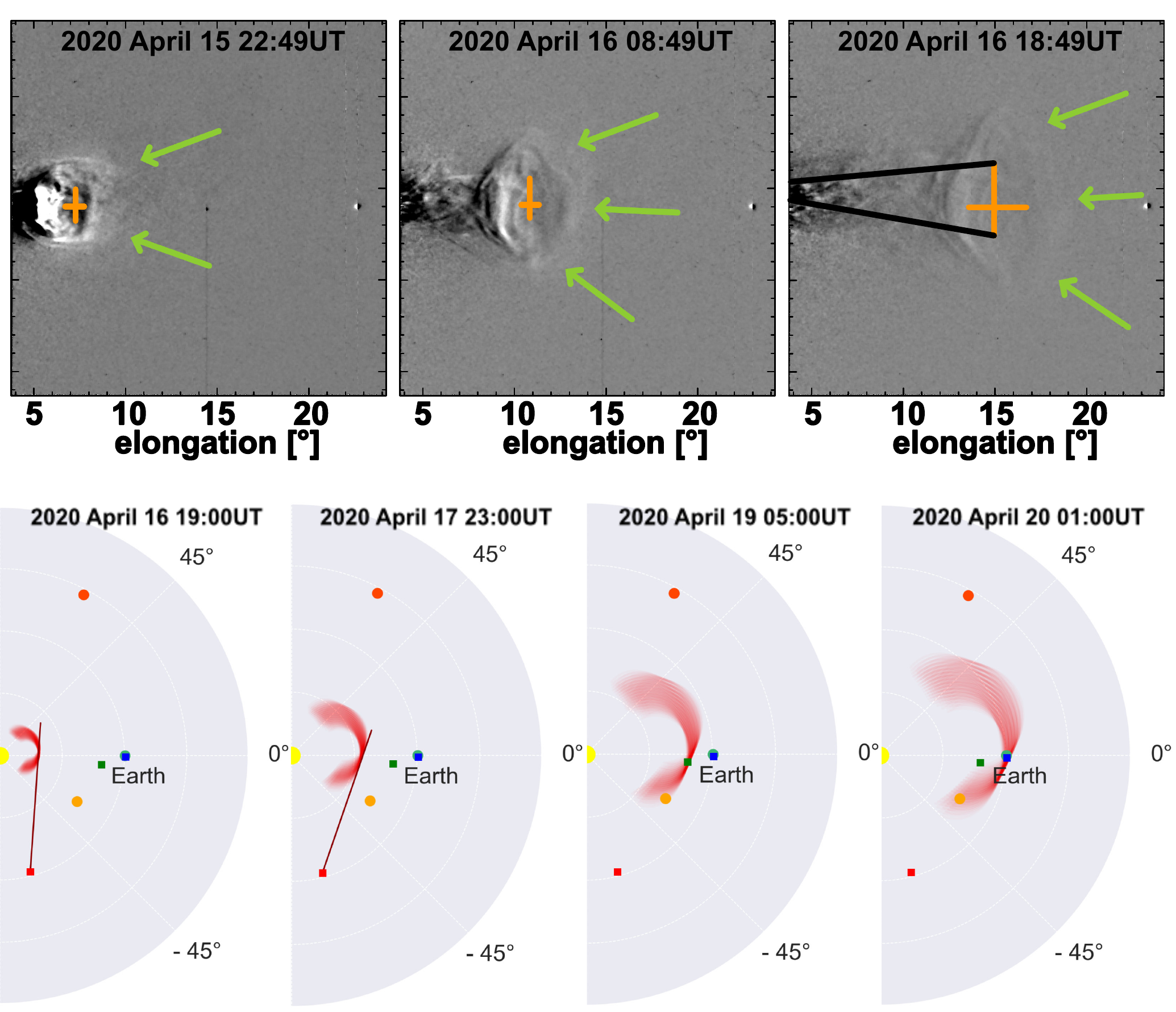}
\caption{Evolution of the CME in STEREO-A HI1. Upper three panels: Running difference images of the CME, with green arrows indicating the CME leading edge. The ecliptic plane corresponds to the horizontal centre of the image, and the elongation angles marked on the x-axis pertain to the ecliptic. The orange crosses mark the extent of the cavity or void, the black lines indicate the angular width of the flux rope. Lower four panels: The propagation of the CME leading edge in the ecliptic plane modelled with an ELEvoHI ensemble simulation, shown for different times, up to the impact of the CME at Solar Orbiter (green square) and Earth. The red tangent corresponds to the elongation measurements from HI.
\label{fig:2_hi} }
\end{figure*}

\textbf{Figure \ref{fig:2_hi}} shows the evolution of the CME in STEREO-A HI1 images. The first HI1 observation of this CME is at 2020 April 15 20:49~UT, when the leading edge is already at around 10\dg elongation. Due to a data gap, no earlier HI1 observations from this event are available. The leading edge of the CME in HI1 images is relatively faint compared to the core following the cavity, which makes accurate tracking of the leading edge of the density structure quite challenging. At 2020 April 16 18:09~UT, the leading edge enters the HI2 field of view at 18.8\dg elongation. Within HI2, the Milky Way is the most prominent feature hindering the tracking of the faint CME front. However, the flattening of the CME cross-section during evolution is perfectly captured by HI1 and enables measurement of the evolution of the aspect ratio of the cross-section for about 22~hours.

The density pile up at the back of the flux rope from the faster solar wind stream helps in increasing the density of the shell wrapping around the flux rope, and thus highlighting its shape in HI images. The CME is clearly viewed as edge-on by HI \citep[e.g.][]{thernisien_2006,wood2009empirical}, which means that the observer looks along the flux rope axis direction. We note that this already gives a seldom made, direct and consistent connection between the CME shape observed in HI observations and in situ observations \citep[cf.][]{mostl2009linking}, where the SEN flux rope type exhibits a low inclination to the solar equatorial plane, consistent with the edge-on view. 

We ran ensembles of the ELEvoHI model \citep{rollett2016,amerstorfer2018} based on the tracked elongation values of the CME front in the ecliptic plane in HI1 and HI2 images. ELEvoHI models the evolution of the CME leading edge in the ecliptic plane, which is in this case the CME shock, as an ellipse that reacts to the ambient solar wind via drag forces \citep{vrsnak_2013}. ELEvoHI was set up using the Ecliptic cut Angles from the graduated cylindrical shell (GCS) model for ELEvoHI tool \citep[EAGEL;][]{hinterreiter2021elevohi}, an ambient solar wind speed range of $425 \pm 200$ km~s$^{-1}$ \citep[for more information on that approach see][]{amerstorfer2020} and a total number of ensemble members of 220. The number of ensemble members comes from running all possible combinations of the range of input parameters (propagation direction, half-width, and inverse ellipse ratio) and the step size used. This results in a median arrival time error between the model results and the observations of $1.8\pm 1.9$ hours (Solar Orbiter) and $-3.3 \pm 1.9$ hours (Earth), demonstrating the unambiguous connection between the imaging and in situ observations. The consistency in modelled arrival time at both spacecraft also demonstrates that no large local distortions of the shock front from the ellipse shape in the ecliptic plane are needed to adequately model the CME propagation between Solar Orbiter and Earth.

A comparison to the results of the graduated cylindrical shell \citep[GCS;][]{thernisien_2006} model by \citet[in revision at A\&A,][]{vonforstner2020radial} also gives a good overall consistency of the HI results with the HEEQ longitude of $18 \pm 6$\dg given by GCS fitting, which used both STEREO COR2 and LASCO C2 coronagraphs. Limiting the ELEvoHI ensemble to the members with an exact agreement of modelled and observed arrival time at Earth, that is $\pm 0.0$ hours, results in an HEEQ longitude of $5 \pm 2$\dg and an angular width (within the ecliptic) of $70 \pm 10$\dg given by the ELEvoHI model. These results are broadly consistent with the shock normal at Wind, calculated by \citet{kilpua2020energetic} using the mixed mode method which resulted in a HEEQ longitude of -1.8\dg. The shock normal points radially away from the Sun and is therefore consistent with a central impact of the shock at Earth.

\subsection{Flux rope modelling} \label{sec:fr_modeling}

In this study, we use three variations of magnetic flux rope models to interpret the large-scale structure of the ICME from the in situ magnetic field profile. As the three models make different assumptions about the flux rope morphology, comparing their correspondence with the observations will enable a greater understanding of the physical processes affecting the ICME evolution.

The first two models are variants of the kinematically-distorted flux rope (KFR) technique \citep{owens_kinematically_2006}. Here, the flux rope is assumed to have a circular cross-section close to the Sun, thus beginning life as a force-free, constant-$\alpha$ solution \citep{lundquist_magnetostatic_1950}. Initial CME shapes can vary due to factors such as the underlying magnetic field geometry \citep{krall2006flux}, however, the assumed circular cross-section is similar to that observed in STEREO COR1-A images of the CME (not shown). The cross-section is then kinematically distorted by radial propagation at the ambient solar wind speed and cross-sectional expansion of the flux rope from internal pressure. This leads to a `pancaking' of the flux rope cross-section \citep{riley_kinematic_2004} that is sometimes inferred from HI observations \citep{savani2010observational}. Two forms of this KFR model are considered: a KFR in a uniform solar wind \citep{owens_kinematically_2006} and a KFR in a solar-minimum-like solar wind, wherein slow solar wind occurs at low latitude and fast wind at high latitude \citep{owens_magnetic_2006}.

For the KFR models, global parameters, such as the flux rope helicity and axis orientation, are determined by minimising the mean squared error (MSE) between the observed and model magnetic field components for a range of model solutions. The start and end boundaries can be free parameters in this minimisation process, but that significantly increases parameter space and the possibility of finding only a local minimum. Instead, here we identify (by eye) discontinuities in the observed magnetic field profile and assess which boundaries produce the maximum eigenvalue ratios in the variance directions \citep[see ][ for more info]{owens2008}. This likely selects only the portion of the flux rope which is unperturbed by strong interaction with ambient solar wind. 

The third model we apply is the 3DCORE flux rope modelling technique \citep[][]{weiss2021analysis}. It uses a 3D torus-like structure, with an embedded Gold-Hoyle like magnetic field that is based on an analytical solution derived by \citet{vandas2018magnetic}. The torus-like structure is anchored on the surface of the Sun and expands self-similarly during its propagation throughout the heliosphere. `Pancaking' or flattening of the cross-section is approximated by using an elliptical cross-section with an aspect-ratio that is either set as constant or a free fitting parameter. Here, we use a value of the aspect ratio that is fixed using the HI observations as a constraint. The flux rope expansion is modelled using a simple power law where the radius of the cross-section expands as $r^{1.14}$ and the magnetic field decreases as $r^{-1.64}$ \citep{leitner2007consequences}. The model parameters are inferred using an approximate Bayesian computation (ABC) algorithm that generates an ensemble of solutions with the root-mean-square error (RMSE) serving as a summary statistic. The fitting is performed on a variable interval within two fixed boundaries which are chosen by manual inspection. In order to compare 3DCORE with the two other models these boundaries were chosen to be very similar. For a more detailed description of the model and the fitting procedure we refer the reader to \citet[]{weiss2021analysis}.

For KFR, the model time series is generated by time evolving the flux rope structure past a fixed point in the heliosphere, approximating a spacecraft encounter with a magnetic cloud. For 3DCORE, the simulation includes a moving observer with the CME propagating within an inertial reference frame. As both KFR and 3DCORE models include flux rope expansion, this leads to an asymmetric magnetic field time series.

\section{Results}\label{sec:results}

\subsection{Radial evolution} \label{sec:radial}

The small longitudinal (within 5\dgg) and latitudinal separation  between Solar Orbiter, Wind, and BepiColombo during the passage of the CME provides an opportunity to study the radial evolution of the CME using in situ measurements of the magnetic field. However, caution must still applied when interpreting results as significant differences in CME magnetic flux rope properties have previously been observed over small longitudinal separations \citep{kilpua2011multipoint, winslow2016longitudinal, davies2020radial} and therefore differences in observations and properties may also be due to spatial variation of the CME. Similarly to the KFR and 3DCORE models, this analysis focuses on the section of the flux rope which is unperturbed (the UMFR) by the interaction with the following higher speed stream and the physical processes occurring towards the leading edge \citep[][in prep.]{kilpua2020energetic}.

\begin{figure}[h]
\includegraphics[width=1.0\linewidth]{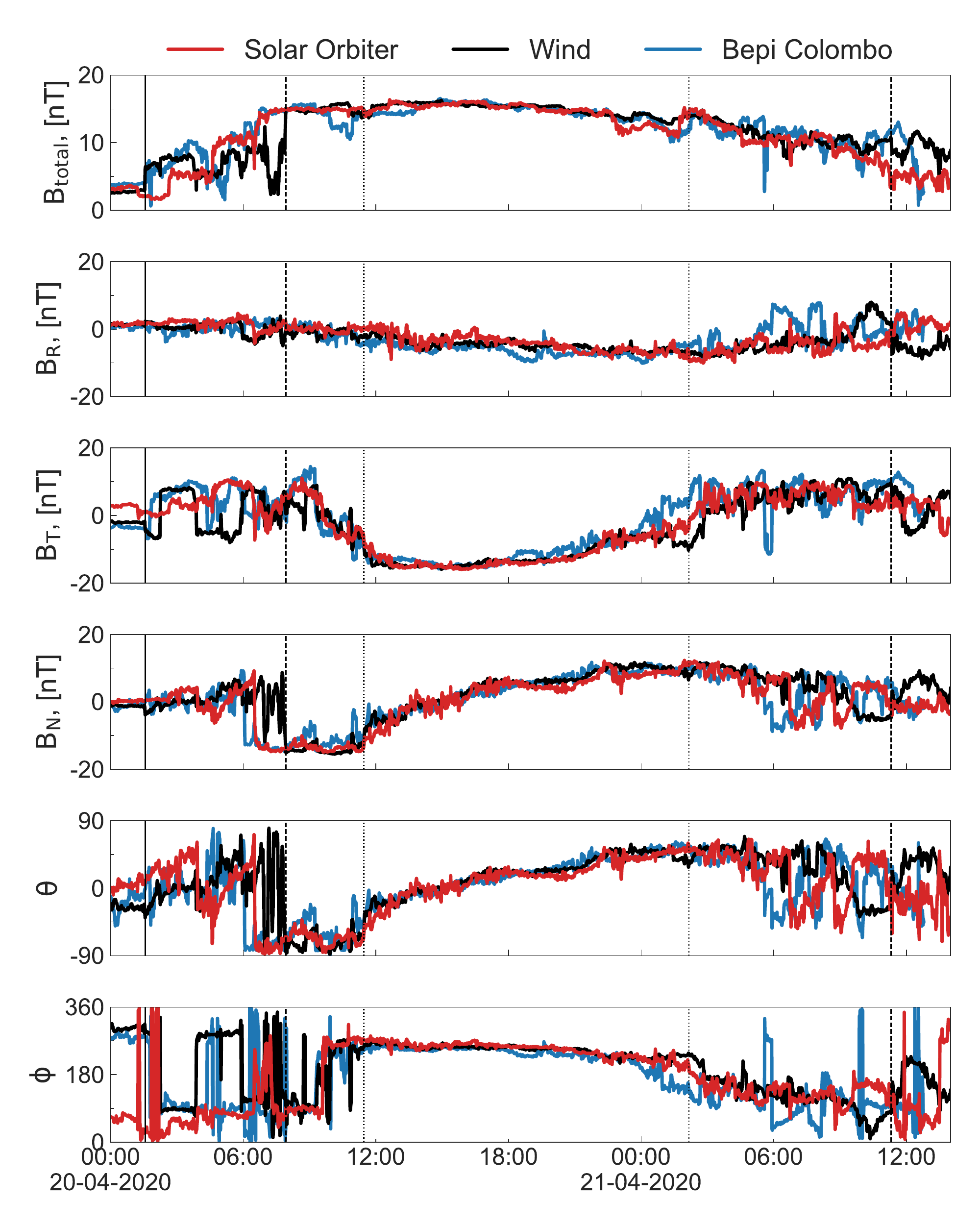}
\caption{Magnetic field measurements at Wind (black), overlaid with the scaled magnetic field measurements of Solar Orbiter (red) and BepiColombo (blue). The vertical solid line indicates the time the shock front was measured at Wind, and the dashed lines constrain the flux rope. The vertical dotted lines within constrain the UMFR, where the magnetic field measurements are well correlated between the three spacecraft.  \label{fig:2_radialevolution}}
\end{figure}

\textbf{Figure \ref{fig:2_radialevolution}} compares the magnetic field profiles at Solar Orbiter, Wind, and BepiColombo. The UMFR has been selected by considering matching features towards the leading and trailing edges in the magnetic field data from the three spacecraft. We compare the durations between different features at each spacecraft to scale the times of the magnetic field time series and maximise the Pearson correlation coefficients. For scaled Solar Orbiter and Wind data the correlation coefficient was found to be 0.87, and for scaled BepiColombo and Wind data the correlation coefficient was found to be 0.80. The start and end times of the identified UMFR are given in Table~\ref{table:1}. Comparing the UMFR duration at Solar Orbiter and BepiColombo to that of Wind gives a scaling factor of 0.998 and 0.901, respectively. The scaled times for each spacecraft are aligned with the start of the UMFR, delineated by the first dotted line in Figure \ref{fig:2_radialevolution}. The scaling factor between Solar Orbiter and Wind is close to unity, which would indicate that the UMFR has undergone little to no expansion as it propagated beyond 0.8 AU to L1. 

The magnetic field strengths of the flux rope at Solar Orbiter and BepiColombo have been scaled using a power law relationship found by fitting the observed maximum magnetic field strength against heliocentric distance at each spacecraft. The magnetic field of the UMFR achieved maximum values of $B_{max}$ = 21.2, 16.3, and 16.1 nT and mean values of $B_{mean}$ = 18.9, 14.9, and 14.6 nT at Solar Orbiter, Wind, and BepiColombo, respectively. Considering the location of each spacecraft, we find the relationships of maximum and mean magnetic field strength with increasing heliocentric distance, $r$, to be $B_{max} \propto$ $r^{-1.24 \pm 0.50}$ and $B_{mean} \propto$ $r^{-1.12 \pm 0.14}$, respectively. Previous studies that have used similar fitting to derive relationships of magnetic field strength with heliocentric distance include \citet{leitner2007consequences, gulisano_global_2010, winslow2015interplanetary}, and \citet{good2019self}. Both \citet{leitner2007consequences} and \citet{gulisano_global_2010} used Helios 1 and 2 observations; \citet{gulisano_global_2010} considered the mean magnetic field to obtain $\langle B \rangle \propto r^{-1.85 \pm 0.07}$ and \citet{leitner2007consequences} fitted magnetic field observations to obtain an axial field relationship of $B_0 \propto r^{-1.64 \pm 0.40}$ for CMEs observed in the inner heliosphere. \citet{good2019self} considered 18 CMEs observed by radially aligned spacecraft including MESSENGER, Venus Express, STEREO A/B, and Wind to find the axial field strength relationship, $B_0 \propto r^{-1.76 \pm 0.04}$, and similarly, \citet{winslow2015interplanetary} considered CMEs observed by MESSENGER and ACE to obtain a relationship of $\langle B \rangle \propto r^{-1.95 \pm 0.19}$. Neither the $B_{max}$ nor $B_{mean}$ relationship calculated in this study agree with previous relationships obtained suggesting that this event is an outlier to overall trends, perhaps due to a more complex interaction with the solar wind and distortion of the flux rope and the possible spatial variation of the magnetic flux rope over the small longitudinal/latitudinal separation between spacecraft. 

For a self-similarly expanding cylindrical force-free magnetic flux rope, one would expect the axial field to decrease as $r^{-2}$ and the poloidal field to decrease with distance as $r^{-1}$ \citep{farrugia_study_1993}. Calculating the RMSE between the scaled Solar Orbiter and Wind time series over the UMFR interval for the ideal axial and poloidal field relationships results in errors of 1.98 and 1.97, respectively. By minimising the RMSE between the time series, the axial field scales as $r^{-1.52}$ and the poloidal field scales as $r^{-0.75}$, with RMSE values of 1.75 and 1.92, respectively. This result implies that the expansion of the CME is not self similar nor cylindrically symmetric, and that the twist distribution changes as the CME propagates.

Using the solar wind speed time series at Wind, we can assess the local expansion of the CME at 1 AU and compare this to the global expansion of the CME. The dimensionless expansion parameter \citep[for more detail, see][]{demoulin_causes_2009, gulisano_global_2010} given by Equation \ref{eq:expansion} provides a measure of local expansion, typically around 0.8 for non-perturbed magnetic clouds \citep{gulisano_global_2010}. The speed at the mid-point of the UMFR, or the cruise velocity, $V_c$ is measured to be 343 km~s$^{-1}$ and the rate of expansion, $\Delta V/\Delta t$, or in this case, $\Delta V_{UMFR}/\Delta_{UMFR} = 7.5\times 10^{-4}$ km~s$^{-2}$. The heliocentric distance, $d$, of Wind at the mid-point of the flux rope is 0.996~AU. 

\begin{equation} \label{eq:expansion}
    \zeta = \dfrac{\Delta V}{\Delta t} \dfrac{d}{V_c^2}.
\end{equation}

\citet{gulisano_global_2010} found that the dimensionless expansion parameter is independent of radial distance and therefore could be used to relate to the global expansion of CMEs with the magnetic field strength. The magnetic field strength would be expected to decrease as $r^{-2\zeta}$ and radial size expected to increase as $r^\zeta$. For the UMFR at Wind, we find that $\zeta$ = 0.95, and therefore would expect the magnetic field strength to decrease with heliocentric distance as $r^{-1.91}$ and the size of the UMFR to scale as $r^{0.95}$. 

The expansion velocity at Wind, calculated as the difference between the velocities at the front and the back of the UMFR is $\Delta V_{UMFR}$ = 39.8 km~s$^{-1}$, and therefore assuming the expansion speed remained constant between Solar Orbiter and Wind, the radial diameter of the flux rope would be expected to increase by 0.02 AU or as $r^{0.87}$. The observed relationship for the radial expansion agrees well with the expected relationship using the dimensionless expansion parameter, however, the magnetic field strength relationships are not in agreement ($r^{-1.91}$ using the dimensionless expansion parameter, in comparison to $r^{-1.12}$/$r^{-1.24}$ found using the observed mean/maximum magnetic field). This discrepancy may be due to compression of the flux rope as a whole by the higher speed stream following the CME.

\subsection{Non-radial evolution} \label{sec:non_radial}

\begin{figure}[h]
\includegraphics[width=1.0\linewidth]{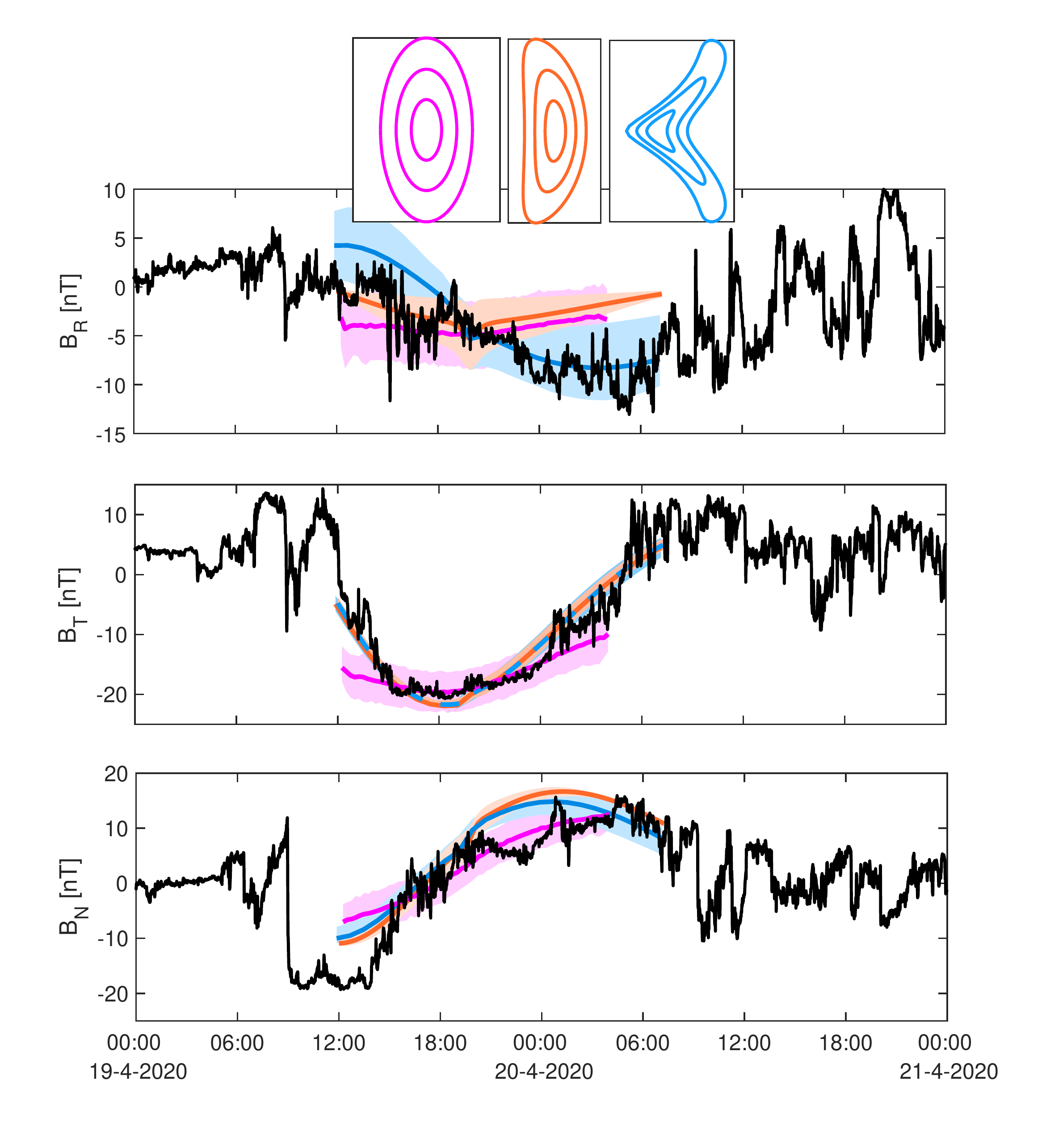}
\caption{Solar Orbiter magnetic field data during the ICME event, overlaid with the three flux rope model time series. The observed 1-minute cadence magnetic field is shown in black, with flux rope model profiles shown in colour. Pink: The 3DCORE fit, which assumes an elliptical flux rope cross-section. Orange: The kinematically-distorted flux rope (KFR) model, assuming a uniform solar wind. Blue: the KFR model assuming a solar-minimum-like latitudinal solar wind speed profile.  \label{fig:4_kfr_3dcore} }
\end{figure}

\begin{figure}[h]
\includegraphics[trim=265px 250px 900px 125px, clip,width=.95\linewidth]{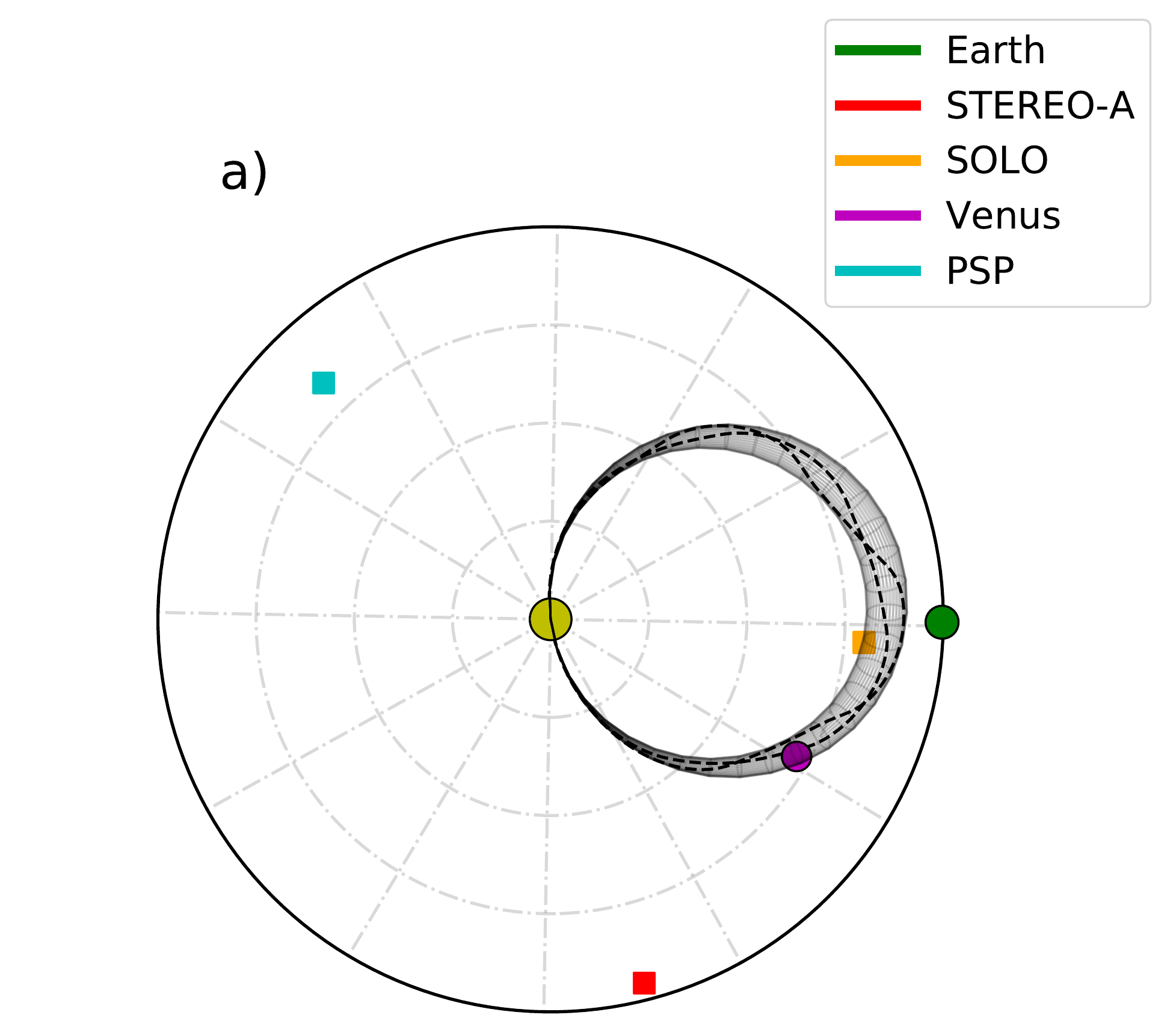}
\includegraphics[trim=1000px 400px 165px 300px, clip,width=.95\linewidth]{3dcore_3d.pdf}
\caption{Three-dimensional reconstruction of the event using a single representative parameter combination from the 3DCORE ensemble solution. \textit{Top panel a)}: top-down view of the 3DCORE geometry within the solar system up to 1~AU with two embedded magnetic field lines (dashed). \textit{Bottom panel b)}: side view of the 3DCORE geometry which shows the chosen cross-section aspect-ratio of $2$.
\label{fig:5_3dcore3d}}
\end{figure}

We now discuss the non-radial evolution of the CME flux rope, first by fits of three different models to the Solar Orbiter magnetic field data, and secondly by considering the STEREO HI observations to demonstrate the amount of flattening of the cross-section for this CME flux rope.

\textbf{Figure \ref{fig:4_kfr_3dcore}} shows the three model solutions for the flux rope cross-section in comparison to the magnetic field time series at Solar Orbiter, in RTN coordinates. In all cases, the global flux rope parameters are reasonably similar, with an axial field lying very close to the $-T$ direction, a left-handed rotation of the flux rope magnetic field about the axis, and a spacecraft encounter close to the axis. 

The cross-sectional shapes of the uniform-wind KFR and 3DCORE flux ropes, shown at the top of Figure~\ref{fig:4_kfr_3dcore} in orange and pink respectively, are generally quite similar, particularly close to the flux rope axis. For this reason, these two models show the same overall variations in the magnetic field components. Firstly, the $B_N$ component shows a strong bipolar signature, passing through zero at the closest approach to the flux rope axis (around Apr 19 18:00 UT). Qualitatively, this is in agreement with the observations. Secondly, the $B_T$ variations are unipolar with the maximum magnitude at the closest approach to the axis. Again, this is in qualitative agreement with the observed variation. Thirdly, the $B_R$ component is significantly smaller in magnitude than $B_T$ and $B_N$, and shows a weak unipolar variation. This is in qualitative disagreement with the observations, which show a dipolar variation in opposition to that of $B_N$. As the uniform-wind KFR assumes cylindrical symmetry and the 3DCORE model allows for a curved flux rope axis, the relative agreement between the two models suggests there is relatively little contribution from axial curvature for this encounter. This is in agreement with an approximately perpendicular spacecraft intersection with the flux rope axis, and the closest approach of the spacecraft being near to the axis \citep{owens_implications_2012}.

The KFR in the solar-minimum-like solar wind shows a very different flux rope cross-section, with a concave-outward profile resulting from faster solar wind at high latitude than the equator. Perhaps non-intuitively, there is little change to the $B_T$ and $B_N$ components for this model compared with the uniform-wind KFR. The most notable difference is to the $B_R$ component, where a bipolar variation is produced, in qualitative agreement with the observed $B_R$. Thus, there is evidence for flux rope distortion by the ambient solar wind.

\begin{figure}[h]
\includegraphics[width=1\linewidth]{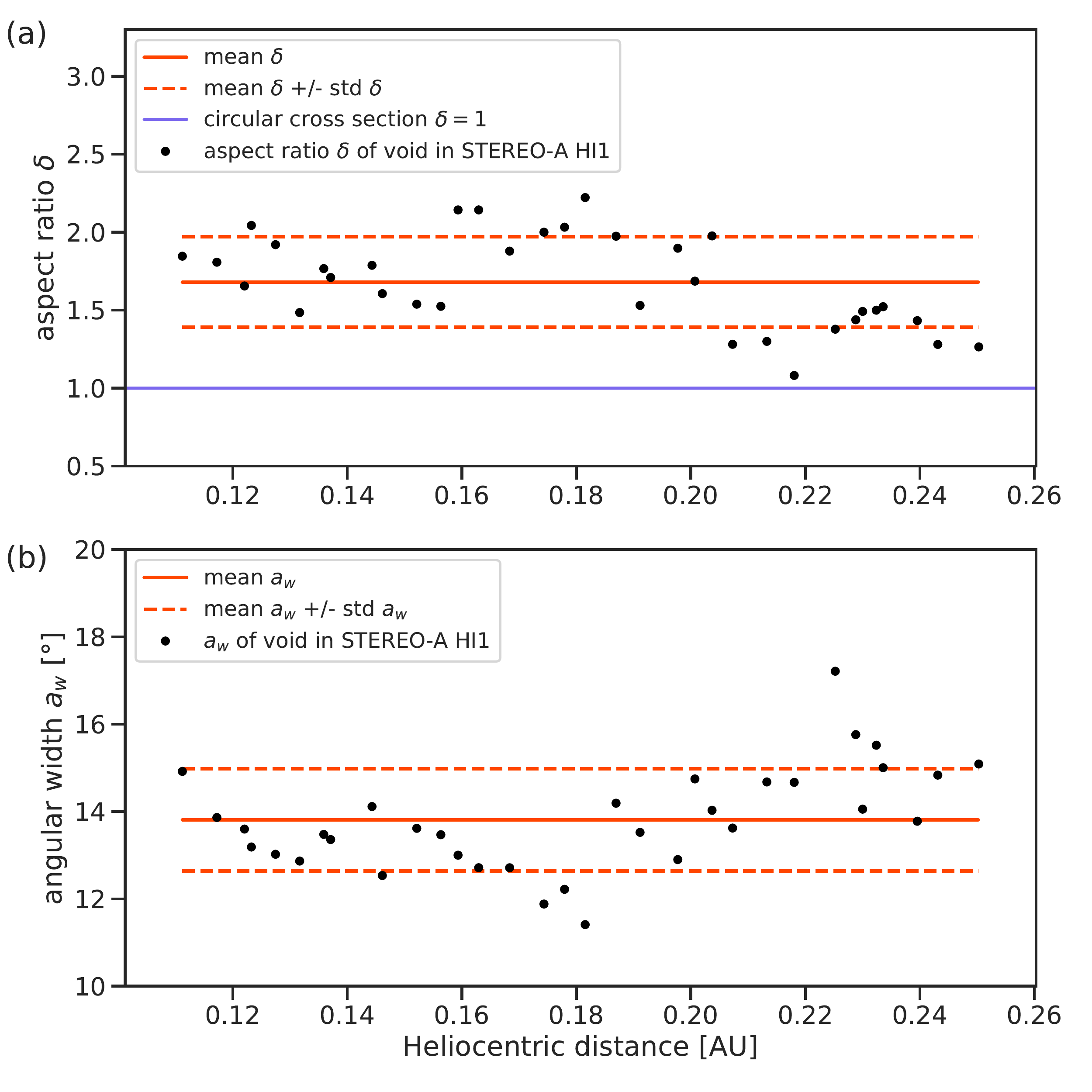}
\caption{Measurements of the aspect ratio $\delta$ and angular width $a_w$ of the CME void in STEREO-A HI1 images, plotted against heliocentric distance. (a) Observed $\delta$ (black dots), mean and standard deviation (red solid and dashed lines), and a blue line at $\delta=1$ indicates the value for a circular cross--section. (b) Angular width $a_w$ derived from STEREO-A HI1 images (black dots), with mean and standard deviation as solid and dashed red lines.\label{fig:6_aspect_aw} }
\end{figure}

\textbf{Figure \ref{fig:5_3dcore3d}} shows a 3D visualisation of a single flux rope fitting solution generated by applying the 3DCORE model to the Solar Orbiter MAG data with a Bayesian minimisation algorithm \citep[][]{weiss2021analysis}. The representative solution is drawn from the full ensemble by a naive attempt at estimating the mode of the multi-dimensional posterior distribution.
The agreement with the Solar Orbiter magnetic field profile is demonstrated by the pink fit in Figure~\ref{fig:4_kfr_3dcore} which shows the generated mean magnetic field profile (solid line) and the 95\% confidence interval that is computed from the ensemble (colored area). 

The main results of this analysis are a direction for the flux rope axis of $13 \pm 5$\dg longitude (HEEQ), $-5 \pm 5$\dg latitude (HEEQ), and a $349 \pm 13$\dg inclination of the axis, with the inclination measured from 0\dg (east) towards 90\dg (south). This means the flux rope axis at Solar Orbiter is elevated roughly 11\dg to the solar equatorial plane, with the eastern part of the flux rope as seen from Earth slightly above this plane, and the western part below it. However, the main result is that the flux rope axis lies almost in the solar equatorial plane. The model flux rope radial diameter at the torus apex is $D_{1AU}=$0.114~AU at 1~AU, and the axial field strength at 1~AU is $B_{1AU}=14.3 \pm 0.9$~nT.

The full number of magnetic field line turns is $\tau=-3.7 \pm 0.6$ over the full torus ($\tau < 0$ means negative chirality), for which the magnetic field structure is assumed as a uniform-twist field in the model (for details, see Section \ref{sec:fr_modeling}). However, the assumption that the flux rope structure extends over the full torus is unlikely to be the case \citep[e.g.][]{owens2016legs}. We propagated the model that resulted from a best fit to the Solar Orbiter data to the Earth L1 point, where, in comparison to Wind data, an inconsistency in the timing is seen, but otherwise the modelled profiles of the magnetic field components match the Wind observations well \citep[see][revised at A\&A]{vonforstner2020radial}. The results of the 3DCORE fitting at Solar Orbiter, Wind, and BepiColombo are also presented by \citet[in prep.][]{weiss2021triple}.

We now proceed to show how we determined the fixed value for the aspect ratio $\delta_{3DCORE}=2.0$ that was used for the fitting procedure, which would otherwise be a free parameter, to show how the HI1 observations can constrain flux rope model parameters in this event.

\textbf{Figure \ref{fig:6_aspect_aw}a} shows measurements of the flux rope aspect ratio by manual selection, in each STEREO-A HI1 image, of the horizontal and vertical extent of the void (see Fig. \ref{fig:2_hi}), which represents the flux rope \citep{davis2009stereoscopic,mostl2009linking} as it has low density and thus low intensity in HI images. We are able to do this because the CME is seen edge-on, thus the HI1 camera looks along the flux rope axis and we see the cross-section of the flux rope directly in HI1 data. The elongation of each observed value in HI1 was converted to a heliocentric distance by use of the ELEvoHI model. For ELEvoHI, values for the longitude to the observer of 82\dgg, an angular width of 70\dgg, and an inverse aspect ratio of the ellipse front shape of $f=0.7$ were taken \citep[see][]{amerstorfer2018,amerstorfer2020}. After converting to distance, we see in Fig.~\ref{fig:6_aspect_aw}a that the HI1 observations covered 0.11 to 0.25 AU. As there is no trend observed for $\delta$ in Fig. \ref{fig:6_aspect_aw}a, we simply derive a mean value of $<\delta_{obs}>=1.68 \pm  0.29$ valid for that distance range.

Given that these direct measurements of the aspect ratio $\delta$ are constrained to distances < 0.25 AU, we shall make another estimate for this parameter that is likely more appropriate further out at 1 AU, as we include the observed radial size for the UMFR observed in situ in the calculation.

\textbf{Figure \ref{fig:6_aspect_aw}b} presents measurements of the angular width $a_w$ (see Fig.~\ref{fig:2_hi}, upper right panel) of the CME void in STEREO-A HI1 images. Again, there does not seem to be a consistent trend with increasing radial distance and we simply take a mean value of $<a_{w}>=13.8 \pm 1.2$\dgg. We note that a constant angular width means that the CME expands self-similarly in the non-radial direction. Assuming $a_w$ stays constant out to 1~AU, we calculate how the extent in angular width would map to a distance along the circumference of a circle with a radius of 1~AU: $d_{latitude}=2 \pi/(360/a_w)= 0.261$~AU. Dividing $d_{latitude}$ by the UMFR radial size of $d_{UMFR}=0.1209$~AU observed at Wind, we find a $\delta_{obs;a_w}= 1.99 \pm 0.17$, with the error arising from the standard deviation in the angular width measurement. This value is quite similar to the previous value determined directly from the HI1 measurements, but it may be more appropriate for 1~AU given that it includes in situ information on the UMFR radial size. Thus we arrive at a value for the aspect ratio $\delta= 2.0$, which we used in the 3DCORE ensemble runs (as previously described).

\subsection{Dst prediction} \label{sec:Dst}

An ICME observed near the Sun--Earth line at $<1$~AU provides the opportunity to study the efficacy of using a spacecraft at such a location as an upstream solar wind monitor to predict the upcoming geomagnetic effects \citep{lindsay1999dst, kubicka2016dst}. We use the PREDSTORM model \citep{Bailey2020} to conduct an analysis of the $Dst$ that would have been forecast at Earth using Solar Orbiter data. This would have provided a lead time of about 20~hours, being the time for the shock to reach Earth after it impacted Solar Orbiter. The method for calculating the $Dst$ index for geomagnetic activity from the solar wind time series follows the same procedure used in \citet{Bailey2020}, partly based on the method from \citet{Temerin2006}. 

\begin{figure}[h]
\includegraphics[width=\linewidth]{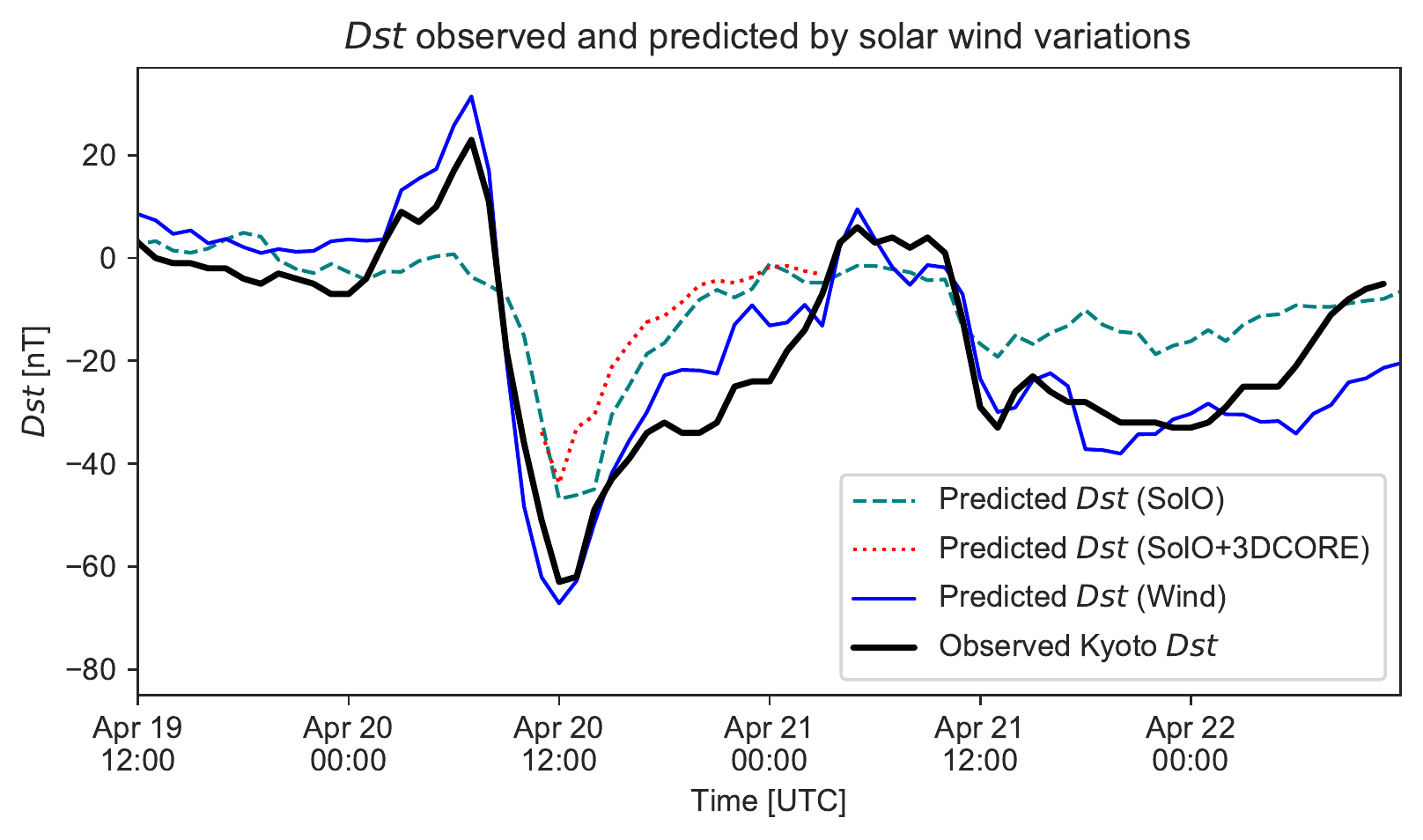}
\caption{\label{fig:dst} Prediction of $Dst$ at Earth using Solar Orbiter data mapped to L1 under the assumption of a constant CME speed (339~km~s$^{-1}$). The solid black line shows the observed Kyoto $Dst$, the solid blue line the $Dst$ predicted by the model using Wind data, while the dashed green line shows the $Dst$ that would have been predicted using a forecast from Solar Orbiter. The dotted red line shows the results for the flux rope magnetic field taken from 3DCORE, which were used to predict the Dst resulting from the second half of the flux rope observed by Solar Orbiter.}
\end{figure}

\textbf{Figure \ref{fig:dst}} shows the $Dst$ predicted using Solar Orbiter data mapped to L1 as well as the Wind L1 data for comparison. The hindcast using Solar Orbiter data is done under the assumption that there is no L1 data to work with, and apart from Solar Orbiter MAG data, only the CME speed derived from HI images is used. The Solar Orbiter magnetic field data is first mapped to L1 with a constant speed and converted from RTN to GSM coordinates for the $Dst$ calculation. Due to the lack of plasma measurements at Solar Orbiter during this time period (including solar wind speed), the value taken was the propagation speed of the CME from HI images with the simpler SSEF30 modelling (339~km~s$^{-1}$) \citep{davies2012self}. Ideally the data would be mapped using the varying speed across the ICME. A constant proton density of 5~cm$^{-3}$ was used throughout the ICME interval to compute the $Dst$ index.

The output from the 3DCORE modelling of the flux rope was also included to demonstrate a forecast from only partial measurements of the flux rope. As can be seen in Fig. \ref{fig:dst}, the $Dst$ forecast from Solar Orbiter compares well with the observed Kyoto $Dst$, although the initial climb and the eventual depth of the trough are both underestimated. The observed $Dst$ minimum of $-60$~nT is, nevertheless, still closely approximated by the modelled minimum of $-47$~nT. The timing of the $Dst$ minimum,  despite a lead time of 20~hours and only an estimate of the solar wind expansion speed to use, is off by 0.5--1~hour. The differences in shape likely result from the use of constant speed and density values (i.e. no defined shock front or sheath region), and could be approximated better if the solar wind structures associated with a CME were accounted for, even if only artificially.

\section{Conclusions}\label{sec:conclusions}

The CME launched on the 2020 April 15 was the first CME to be detected in situ by Solar Orbiter on April 19. The CME was also observed in situ by both the Wind and BepiColombo spacecraft on April 20, whilst BepiColombo was longitudinally separated from the Earth by just 1\dgg. The radial separation of 0.19~AU between Solar Orbiter and Wind/BepiColombo and a longitudinal separation of less than 5\dg has provided an excellent opportunity to study the radial evolution of the CME.  

STEREO-A was located at an exceptionally well suited viewpoint (75.1\dgg longitude) for heliospheric imaging of an Earth directed CME. The source of the CME was an almost entirely isolated streamer blowout which was pushed from behind by a higher speed stream. This caused a density pile up at back of the flux rope, confirmed by in situ measurements of the density at Wind. This is evidenced in STEREO-A HI1 images by the increased intensity of the density shell wrapping around the flux rope, highlighting the shape of the CME. A clear flattening of the CME as it propagated has been observed in the HI1 images. The CME was viewed edge-on by HI1, thus the camera was aligned along the flux rope axis direction. This gives a direct and consistent connection between the CME shape observed by HI1 with magnetic field instruments in situ, where the SEN flux rope type indicates a low inclination to the solar equatorial plane, consistent with the edge-on view from STEREO-A. The remarkably clear HI1 images of the CME have allowed direct measurements of the flux rope cavity to be made. These measurements have been used to calculate an aspect ratio of 2, further confirming the flattening of the flux rope cross-section observed as the CME propagated from the Sun to 1~AU. 

We investigate the non-radial evolution of the CME flux rope by fitting three flux rope models to the Solar Orbiter magnetic field data. The magnetic flux rope models each make different assumptions about the flux rope morphology to interpret the large-scale structure of the ICME. The 3DCORE model assumes an elliptical cross-section with a fixed aspect-ratio calculated by using the HI1 observations as a constraint. The cross-sectional shape is relatively similar to that of the KFR model in the uniform solar wind and therefore, these two models show similar general variations in their output magnetic field components. Despite the cylindrical symmetry of the uniform-wind KFR model in comparison to the 3DCORE model, the latter of which allows for a curved flux rope axis, the agreement between models suggests that there is relatively little contribution from axial curvature, in agreement with an approximately perpendicular spacecraft intersection with the flux rope axis, and the spacecraft crossing the flux rope close to its axis. The KFR model profile in a latitudinally structured solar-minimum-like solar wind shows the most notable differences in results, where a bipolar variation in the radial magnetic field component is produced. This is in qualitative agreement with the observed radial magnetic field at Solar Orbiter and thus, provides strong evidence for flux rope distortion by the ambient solar wind. 

The radial evolution of the CME has been assessed by scaling and overlaying the magnetic field data from Solar Orbiter, Wind, and BepiColombo. The boundaries of the CME are ambiguous, both towards the leading edge of the flux rope due to ongoing processes and features in the sheath, and towards the trailing edge due to the higher speed stream following the CME. Analysis therefore focuses on the section of the flux rope which is unperturbed. The so-called unperturbed magnetic flux rope, UMFR, has been identified independently in both the KFR model technique where (1) boundaries in the observed magnetic field were identified and assessed to maximise eigenvalue ratios in the variance directions, and (2) in the magnetic field mapping technique, where different durations between magnetic features towards the leading and trailing edge were compared to maximise the Pearson correlation coefficient between spacecraft time series. We find that the magnetic field measurements are well correlated between spacecraft for the UMFR interval. Comparing in situ observations of the magnetic field from the different spacecraft, we find the dependence of the maximum (mean) magnetic field strength with heliocentric distance to decrease as $r^{-1.24}$ ($r^{-1.12}$), in disagreement with previous studies where one would expect the magnetic field to decrease at a faster rate than observed in this study. Further assessment of the axial and poloidal magnetic field strength dependencies suggest that the expansion of the CME is likely not self-similar nor cylindrically symmetric, in agreement with the evidence from the HI1 imagery that the flux rope cross-section was distorted as it propagated due to a more complex interaction with the solar wind. 

The CME propagated close to the Sun-Earth line providing a good opportunity to demonstrate the efficacy of using Solar Orbiter, whilst in such an ideal location, as an upstream monitor to predict the Dst index observed at Earth. The in situ observations of the magnetic field at Solar Orbiter have been mapped using the PREDSTORM model, producing a relatively accurate estimate of the Dst index, with a Dst minimum of $-47$~nT comparable to observations at Earth of -60~nT, despite the model assumption that there was no data available at L1. 

The configuration of Solar Orbiter, STEREO-A and spacecraft near the Earth during the propagation of the CME between 2020 April 15 and April 20 has provided highly useful observations to determine the large-scale shape of the CME using both in situ and remote observations in combination with modelling. This study contributes towards our understanding of the radial and non-radial evolution of CMEs through the inner heliosphere, whilst questioning the traditional picture of CME morphology and flux rope model profiles. Further studies are necessary to determine the commonality of CME flux rope distortion and the effects of CME interaction with varying solar wind structure. As we enter Solar Cycle 25 and solar activity increases, the Solar Orbiter mission, as it approaches the Sun, will contribute to an exciting opportunity for future multi-spacecraft studies utilising both in situ and remote observations of CMEs at various radial distances.

\begin{acknowledgements}

We have benefited from the availability of Solar Orbiter, STEREO, Wind, and BepiColombo data, and thus would like to thank the instrument teams, and the SPDF CDAWeb and Solar Orbiter (\url{http://soar.esac.esa.int/soar/}) data archives for their distribution of data. The Solar Orbiter magnetometer was funded by the UK Space Agency (grant ST/T001062/1). This research was supported by funding from the Science and Technology Facilities Council (STFC) studentship ST/N504336/1 (E.D.) and STFC grants ST/S000361/1 (T.H.) and ST/R000921/1 (M.O.).
C.M., A.J.W, R.L.B, M.A.R., T. A., J.H., M.B. thank the Austrian Science Fund (FWF): P31521-N27, P31659-N27, P31265-N27.
D.H., I.R., H.-U. A. were supported by the German Ministerium für Wirtschaft und Energie and the German Zentrum für Luft- und Raumfahrt under contract 50 QW 1501.
D.B., J.D., R.H. recognise the support of the UK Space Agency for funding STEREO/HI operations in the UK. 

\end{acknowledgements}

\newpage
\bibliographystyle{aa}
\bibliography{final_submission}

\clearpage

\end{document}